\documentstyle[twocolumn,aps,psfig,epsf]{revtex}
\newcommand{\la}{\langle}
\newcommand{\ra}{\rangle}

\begin{document}
\twocolumn[\hsize\textwidth\columnwidth\hsize\csname 
 @twocolumnfalse\endcsname
\title{Universal fluctuations and extreme value statistics }
\author{Kajsa Dahlstedt$^{a,b}$ and Henrik Jeldtoft Jensen$^{a*}$ }
\pagestyle{myheadings}
\address{$^{a}$Department of Mathematics, Imperial College, 180 Queen's Gate, 
London SW7 2BZ, UK \\ 
$^b$Department of Physics, Royal Institute of Technology (KTH), SE-100 44 Stockholm, Sweden
}

\maketitle
\date{\today}
\begin{abstract}
We study the effect of long range algebraic correlations on extreme value
statistics and demonstrate that correlations can 
produce a limit distribution which is indistinguishable
from the ubiquitous Bramwell-Holdsworth-Pinton
distribution. We also consider the square-width fluctuations
of the avalanche signal. We find, as recently predicted
by T. Antal, M. Droz G. Gy{\"o}rgyi and Z. R{\'a}cz
for logarithmic correlated $1/f$ signals, that these fluctuations follow
the Fisher-Tippett-Gumbel distribution from uncorrelated extreme
value statistics.
\end{abstract}
\vskip1pc
]

\section{Introduction}
Three years ago Bramwell, Holdsworth and Pinton (BHP) published the
remarkable discovery that the same functional form that describes
the fluctuation spectrum of the energy injected in an experiment on turbulence also describes the fluctuations in the magnetization of the finite 
size two dimensional XY equilibrium model in the critical region
below the Kosterlitz-Thouless transition temperature \cite{bhp_nature}.
Since three dimensional turbulence and two dimensional
magnetic equilibrium system appear to have very little
in common, BHP made the reasonable suggestion
that the origin of the identical functional form for the fluctuation spectra
should be sought in the one thing the two systems appear to share,
namely, {\em scale invariance}. This suggestion was supported
by the subsequent finding that a long list of scale invariant 
(or nearly scale invariant) non-equilibrium as well 
as equilibrium systems exhibit the same BHP form for the fluctuation
spectrum for certain quantities \cite{b_et_al}. 

Nevertheless,
not all critical systems fluctuates according to the BHP 
spectrum. This was made very explicit by Aji and Goldenfeld \cite{aji},
who proffered the interesting suggestion that the reason the 
two dimensional (2d) XY-model and the driven turbulence experiment
exhibit the same fluctuation spectrum could be that the 2d XY-model
{\em is} the effective model for the turbulence experiment.
Even if this is correct we still lack an explanation of why so many
disparate systems \cite{b_et_al} do exhibit BHP fluctuations.

The BHP functional form is similar to one of the asymptotic
forms for extreme value statistics: the asymptote first
discussed by  Fisher and Tippett \cite{fisher} and often
referred to as the Gumbel distribution \cite{E_V_stat}. The BHP distribution
is, however, not identical to the Fisher-Tippett-Gumbel (FTG)
asymptote. 
Consider $T$ independent and identically distributed  
stochastic variables. Under certain conditions (essentially
exponential tail) the $k$-th largest of the $T$ variables will
be distributed according to the FTG
asymptote with $k$ entering as a parameter. The BHP form
can be thought of as corresponding to the somewhat
uninterpretable case of  $k=\pi/2$ \cite{b_et_al}.

As already alluded to in Ref. \cite{b_et_al} the deviation
between the BHP and the FTG form may be related to correlations.
Our main aim in the present paper is to study this point
in detail. We do that by simulating the so called Sneppen depinning
model \cite{sneppen}. The power spectrum of the Sneppen model
behaves like $1/f^\beta$ with $\beta\simeq 0.5$ for low frequencies
corresponding to a very slow algebraic decay of the autocorrelation
function (see e.g. \cite{SOC_book}).
We use the Sneppen model in the next section to demonstrate that 
extreme value statistics of $T$ strongly correlated exponentially
distributed variables may follow the BHP form. Remarkably, Antal et al. \cite{antal} demonstrated analytically that, at least for a certain class of $1/f$ signals (periodic signals), the width-square fluctuations ($w_2$)
of the signal follow the FTG distribution from extreme value statistics, though 
it is not clear why this should be the case. Inspired by this finding we study
in Section III the width fluctuations of the avalanche signal in the
Sneppen model. We find, contrary to the 1/f case studied by Antal et al.,
that the probability density function for $w_2$ in the Sneppen model is very 
well represented by the FTG distribution even for non-periodic signals. Section 
IV contains a discussion and our conclusions.

\section{Extreme value statistics and the Sneppen model}
To investigate the relationship between extreme value statistics of
correlated variables and the BHP probability density we consider now the 
simple 1+1 dimensional depinning model introduced by Sneppen \cite{sneppen}.
The model is imagined to represent a one dimensional elastic interface moving transverse while acted upon by a set of random pinning forces, See Fig. 1a.

The model,
which is discrete and  very schematic, consists of $L$ sites in the
$x$-direction and infinitely many sites in the $y$-direction. Each square on this 
semi-infinite lattice is assigned a random number, the pinning force, uniformly 
distributed on the interval $[0,1[$. The interface is represented by the set 
$\{(x,y)|y=h(x,t)\}$, where $h(x,t)$ denotes the ``hight'' of the interface 
above the $x$-axis at time $t$ at site $x$ along the $x$-axis. The initial 
configuration is $h(x,0)=0$ for $\forall \; x$. In each time step the interface 
site, $x_s$ with the {\em smallest} pinning force is located and the interface
at this location is moved one step ahead, i.e., $h(x_s,t)\mapsto h(x_s,t)+1$.
This may cause the neighbouring slope $|h(x_s,t)-h(x_s-1,t)|$ (or
$|h(x_s+1,t)-h(x_s,t)|$) to exceed 1, in which case the interface at site
$x_s-1$ is moved one step ahead, i.e. $h(x_s-1,t)\mapsto h(x_s-1,t)+1$
(and similar for the site $x_s+1$ if needed). The update of the nearest 
neighbour sites of $x_s$ may cause the slope on the next nearest sites to
 exceed 1. In which case the interface is moved ahead on these sites. This
procedure is repeated until all slopes satisfy $h(x,t)-h(x-1,t)|\leq 1$
once again. The sequence of operations needed to make all slopes smaller
than or equal to 1 after the update of site $x_s$ is denoted {\em one
time step}.  The number of sites updated during one time step is called
a micro-avalanche. The number of sites being updated during the time step 
$t$ is called the size of the avalanche $s(t)$ and per definition the
duration of each of these avalanches is the same, namely one time step.

\begin{figure}[ht]
\vspace{.75cm}
\centerline{\hspace{-0cm}
\psfig{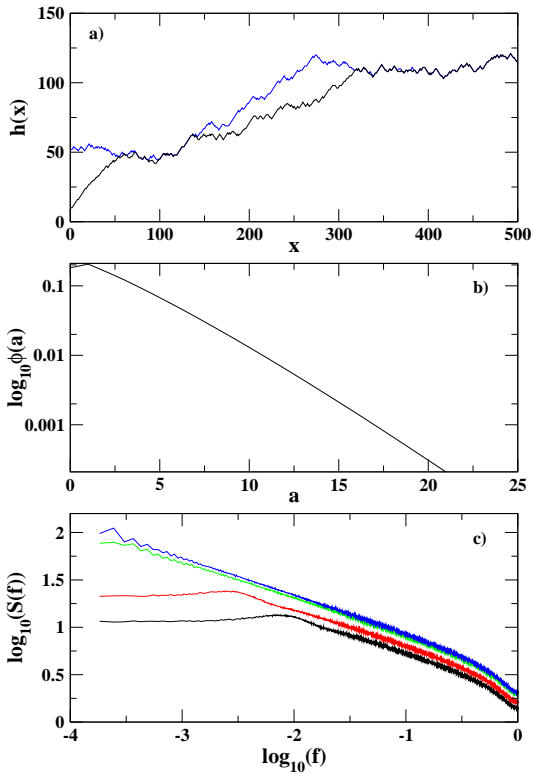}}
\vspace{.75cm}
\caption{
{\bf a)}  Snapshot of the sneppen interface $h(x)$ at two times 
10000 timesteps apart, system size $L=500$. {\bf b)} The PDF of the micro-avalanches in the sneppen model, system size $L=500$. {\bf c)} The power spectrum of micro-avalanches for 4 different system sizes. From bottom to top  $L=50,100,500,1000$. 
} 
\label{1}
\end{figure}

The probability density function (PDF) for the size of the micro-avalanches is 
close to an exponential
as seen in Fig. 1b and first reported in \cite{sneppen}. The individual
micro-avalanches are strongly correlated in time. This is clearly
seen from the power spectrum $S(f)$ of the temporal signal $s(t)$, see
Fig. 1c. The low frequency behaviour of the power spectrum is
approximately  $S(f)\sim 1/\sqrt{f}$, indicative
of slow algebraic decay of the auto-correlation function of
the signal $s(t)\sim 1/\sqrt{t}$ (see e.g. \cite{SOC_book}). 

Let us now imagine that the activity of the model is monitored by some 
devise which has only a limited resolution. This might be modeled by
assuming that the measured signal $\bar{S}_T(\tau)$, rather than 
consisting of the microscopic
instantaneous activity $s(t)$, is given by the sum of activities 
within a time window of a certain size $T$:
\begin{equation}     
\bar{S}_T(\tau) = \sum_{t=\tau}^{\tau+T} s(t).
\label{S_tau}
\end{equation}
This signal was among the set of quantities in Ref. \cite{b_et_al} found
to follow the BHP distribution for a certain range of window sizes $T$.
That is, if $T$ is too small the PDF for $\bar{S}_T(\tau)$ is close to an exponential
distribution. The central limit theorem will, however, cause the PDF for
$\bar{S}_T(\tau)$ to approach a Gaussian when $T$ becomes large enough that correlations
amongst the $s(t)$ entering the sum in Eq. \ref{S_tau} can be neglected.
This deviation can be observed in Fig. 2a. Presumably the PDF for $S(T)$ would
remain close to the BHP form even for $T\rightarrow\infty$ if the correlation
time of the signal was infinite as it would be expected to be in the limit
of infinite system size, see Fig. 1c.

Since, as seen in Fig. 2b, the functional form of the BHP distribution is close 
to the first FTG asymptote for extreme value statistics we proceed to 
investigate  the extreme statistics of the correlated variables $s(t)$ generated by 
the Sneppen model. For the time windows considered for $\bar{S}_T(\tau)$ we define
\begin{equation}
M_T(\tau) = \max\{s(\tau),s(\tau+1), . . .,s(\tau+T)\}.
\label{M_tau}
\end{equation}

In Fig. 2b we show the scaled PDF for $M_T(\tau)$. First we compare the distribution
of the sum and the max in Fig. 2a. We see that the distribution for $M_T(\tau)$
remain very close to the BHP distribution for all considered sampling sizes
$T$, while the distribution for $\bar{S}_T(\tau)$ gradually deviates from the BHP
as $T$ is increased. In Fig. 2b we demonstrate two interesting points. First
that for system size $L=1000$ we are unable to make $T$ so large that the 
distribution for $M_T(\tau)$ deviates form the BHP form. No difference 
between the results for $T=5000$ and $T=10000$ can be detected. 
The size dependence is, however, detectable, as illustrated in the insert to
Fig. 2b. We show here the PDF for smaller system sizes $L=500$ and $L=50$.
The spatial extend of the system $L=50$ is now sufficiently small to destroy the long range temporal correlations as is seen from the fact that the
PDF for $L=50$,  $T=500$ follows the FTG form for uncorrelated extreme value statistics.
We elaborate on the role of the correlations in the main frame of Fig. 2b. 
This figure demonstrates that the deviation between the PDF for $M_T(\tau)$ and the FTG form is indeed caused by the correlations of the Sneppen model. This 
conclusion is reached in the following way. We generate $T$ uncorrelated 
stochastic variables $\chi_i$ all drawn from the PDF for the individual micro-avalanches $s(t)$ (see Fig. 1b). The difference
between
\begin{equation}
M_{uc}=\max\{\chi_1,\chi_2, . . . ,\chi_T\}
\end{equation}
and $M_T(\tau)$ is solely the correlations amongst the primary variables from 
which $M_T(\tau)$ is generated. Hence, given the exponential form of the PDF for
the distribution of the individual micro-avalanches (see Fig. 1b) we expect
the PDF for the {\em uncorrelated} extreme $M_{uc}$ to follow the FTG asymptote for large values of $T$. This is exactly what is found in Fig. 2b. 

\begin{figure}[ht]
\vspace{-0cm}
\centerline{\hspace{-0cm}
\psfig{figure=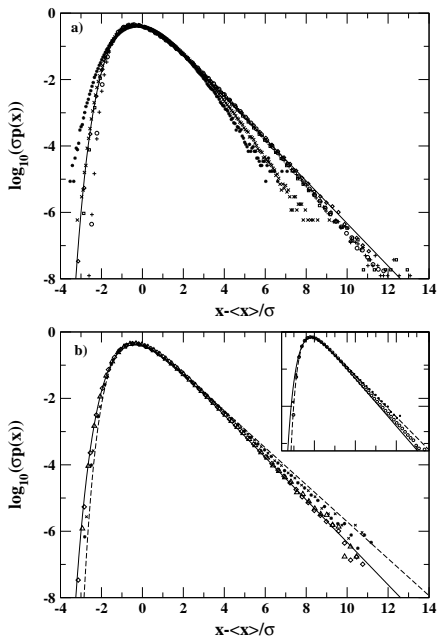,width=7.cm,angle=0}}
\vspace{.5cm}
\caption{
{\bf a)} The PDF of $\bar{S}_T(\tau)$ for $T=50(+)$, $T=500(x)$ and 
$T=500(*)$ and the PDF of $\bar{M}_T(\tau)$ for $T=50 \;(\circ) $, 
$T=500\;(\Box) $ and $T=5000\;(\diamond)$, the solid line is the BHP PDF. 
{\bf b)} The PDF of $M_T(\tau)$ for $T=5000 \;(\diamond)$ and 
$T=10000\;(\triangle)$ and the PDF of the decorrelated maximum 
$M_uc$ for $t=5000\;(*)$ and $T=10000\;(+)$, the solid line is the 
BHP form and the dashed line is the FTG form. {\bf The inset:} The 
PDF of $M_T(\tau)$ for system sizes,  $L=50$, $T=50\;(\circ)$ and 
$T=500\;(\bullet)$.
} 
\label{2}
\end{figure}

\section{The square width and extreme value statistics}
In the present section we analyse the statistics of the fluctuations in the
width of the avalanche signal $s(t)$ generated by
the Sneppen model and of the $1/f$ signal $F(\tau)$ in Eq.
\ref{F_tau} below. Our inspiration is from the beautiful paper
by Antal, Droz, Gy{\"o}rgysi and  R{\'a}cz (ADGR) \cite{antal}. These
authors demonstrated analytically  that at least for a certain class of
$1/f$ signals $h(t)$, the width-square 
\begin{equation} 
w_2(h) = \overline{[h(t)-\bar{h}]^2} 
\label{w_2} 
\end{equation} 
is distributed according to the FTG distribution. In Eq. \ref{w_2}
the over-bar denotes the following time average
\begin{equation} 
\bar{h} = {1\over T}\sum_{t=1}^T h(t).
\end{equation}
The result by ADGR is striking since it is unclear how 
or if extreme value statistics is involved in some effective way
in determening the distribution of $w_2$. 

As a contribution to the investigation of the generality of the
discovery by ADGR we studied $w_2$ for the signal 
$s(t)$. This signal is not $1/f$ but rather 
$1/\sqrt{f}$, and differs in this way from the class studied in 
Ref. \cite{antal}. As see in Fig. 3a $w_2$ for $s(t)$ does
follow the FTG functional form except for very small sample sizes.
This indicates that the ADGR result is more general than their calculation
allows one to conclude. 
ADGR predicted analytically that this should be the case for 
periodic $1/f$ signals and they report that deviations are observed
in simulations of non-periodic $1/f$ signals.

To address the question about deviations from the FTG form
we show in Fig. 3b the fluctuations in $w_2$ for a
$1/f$ signal of the type considered by Antal et al. We
generated the signal as the ``half integral'' of white noise \cite{press}, i.e.,
the signal is obtained as a convolution between white noise and
a propagator $G(t)$ which decays as an inverse square root:
\begin{equation}  
F(\tau) = \lim_{T_*\rightarrow\infty}\sum_{t=\tau-T_*}^\tau  G(\tau-t)\chi(t), 
\label{F_tau} 
\end{equation} 
where $G(x) = 1/\sqrt{x}$. We assume $\chi(t)$ to be uncorrelated
and uniformly distributed on the interval $[-1,1]$ and in our
simulations we truncate the sum in Eq. (\ref{F_tau}) at $T_*= 2^{20}$. 

One can calculate the PDF of the Fourier coefficients of the signal 
of length $T$ in Eq. (\ref{F_tau}) by performing a discrete Fourier analysis  
(note this amounts to assuming a signal of periodicity $T$) 
\begin{equation}
\hat{F}(\omega)= \sum_{\tau=0}^{T} F(\tau)e^{-i\omega\tau}.
\label{FT}
\end{equation}
Averaging over the white noise $\chi(t)$ gives the following PDF for
the Fourier amplitudes
\begin{equation}
p(|\hat{F}(\omega)|^2=R^2)= {3\omega\over \pi T}\exp\left(-{3\over\pi T}\omega R^2\right).
\label{PDF_of_FT}
\end{equation}
which demonstrates explicitly that the signal in Eq. (\ref{F_tau}) is
of the type considered by ADGR \cite{antal}.

We notice that the simulated distribution for $w_2$ in Fig. 3b 
for all sample sizes $T$ deviates from the FTG form predicted by
Antal et al. This was indeed noticed by ADGR (their Fig. 1)
and ascribed to the difference between the assumed periodic boundary condition
needed to perform  analytical calculation of the PDF of $w_2$ and the
non-periodic signal simulated. This problem appears not to arise
in case of the Sneppen model, where Fig. 3a shows that the PDF
for $w_2$ follows the FTG form as soon as $T$ is of order 500 or larger.
We recall that the difference between the signal in
the Sneppen model and the signal in Eq. (\ref{F_tau}) is that
the Sneppen signal is algebraically correlated with an
autocorrelation function that decays like one over the square root
of time whereas the $1/f$ signal in Eq. (\ref{F_tau}) 
decays even more slowly, namely logarithmically. It is likely
that this difference in the range of the correlations is
the cause of the difference between the two signals sensitivity 
to boundary condition.

\begin{figure}[ht]
\vspace{1cm}
\centerline{\hspace{-0cm}
\psfig{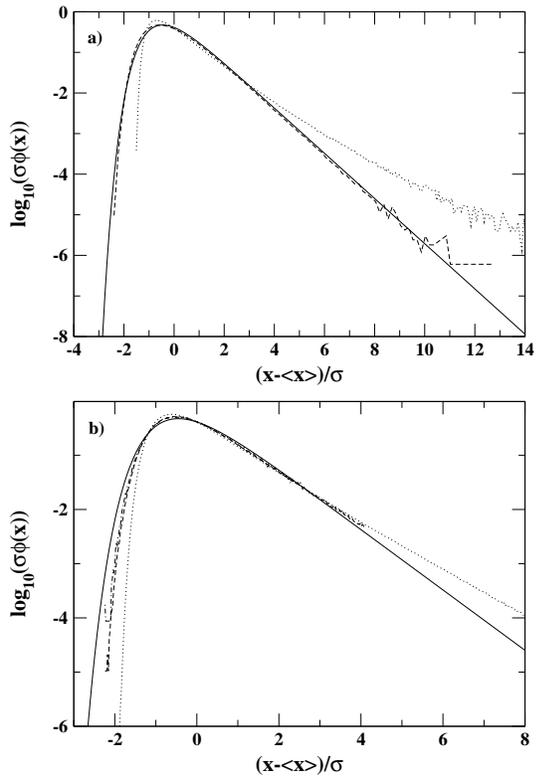}}
\vspace{.75cm}
\caption{
{\bf a)} The PDF of the width-square for the sneppen model, sample sizes $T=50$ (dotted line) and $T=500$ (dashed line), the solid line is the FTG form.
{\bf b)} shows the PDF of the width-square for the 1/f-noise, sample sizes $T=64$ (dotted line), $T=4096$ (dashed line) and $T=32768$(dashed-dotted line), the solid line is the FTG form.
} 
\label{3}
\end{figure}

\section{Discussion and Conclusion}
We have studied the effects of correlations on extreme value statistics.
Our aim has been to relate the Bramwell, Heldsworth and Pinton 
distribution \cite{bhp_nature,b_et_al} to correlated extreme 
statistics. We have achieved this successfully for one specific
system: the Sneppen depinning model.

We found that the extreme value statistics of
the algebraically correlated avalanche signal 
is described by the BHP distribution. Furthermore 
we studied the width-square of the
same signal and found this to be distributed according to
the Fisher-Tippett-Gumbel distribution for {\em uncorrelated} 
extreme value statistics, though no extremes were explicitly involved for
the width-square signal. This finding generalises a recent result by 
Antal, Droz, Gy{\"o}rgyi and R{\'a}cz \cite{antal}.
The BHP distribution and the FTG distribution are
of similar functional form though they differ significantly
for large fluctuations away from the mean, see \cite{BH_long}. 

Extreme value statistics for $1/f$ correlated signals were studied
using renormalisation group (RG) techniques by Carpentier and Le Doussal 
\cite{carpentier}. They find that the correlations make the functional
form change from the exponential decay of the uncorrelated FTG 
to the form $y\exp(-y)$ for fluctuations $y=x-\la x\ra\gg0$ above
the average  (in the case of Maximum statistics). 
Note the $y\exp(-y)$ decay is identical to the asymptotic 
behaviour of the BHP distribution for large deviations above the
average \cite{b_et_al,BH_long}. Hence, the RG calculation 
by Carpentier and Le Doussal is an analytic indication 
of a relationship between the BHP distribution
and extreme value statistics of logarithmically correlated variables.
Our numerical study indicates a relationship in the case
of algebraic correlations, at least in the specific
example of the Sneppen model. 

Signals with hierarchical correlations were considered in a recent preprint
by Dean and Majumdar \cite{dean}. They find that correlations
typically, though not always, alters the super-exponential $\exp(-\exp(y))$ 
found for $y<0$ in the uncorrelated FTG case, as well as in the BHP 
distribution. This result makes it clear that not all correlations
make extreme value statistics follow the BHP distribution.

We conclude that correlated extreme value statistics may produce 
a limit distribution indistinguishable from the Bramwell-Holdsworth-Pinton
distribution. More research is needed in order to determine the generality
of this result.

\section{Acknowledgement}
It is a pleasure to acknowledge stimulating discussion with our friends and colleagues at Imperial, especially K. Christensen, D. Hamon, M. Hall, S. Lise, M. Nicodemi and  G. Pruessner. We are also indebted to G. Pruessner for pointing out Ref. \cite{antal}.
HJJ greatly appreciates the exchange of general ideas concerning universal 
fluctuations with  J. Lopez, P. Holdsworth and S. Bramwell.
KD is supported by KTH.

\vspace{1cm}
\noindent $^*$ Author  to whom correspondence should be addressed. E-mail: 
h.jensen@ic.ac.uk

\end{document}